\documentclass[prb,twocolumn,showpacs]{revtex4}
\usepackage{epsf}
\usepackage{times}
\begin{document}

\title{Shot noise of spin current in ferromagnet-normal-metal
systems}

\author{M. Zareyan$^{1,2}$ and W. Belzig $^{3}$}

\affiliation{ $^{1}$ Institute for Advanced Studies in Basic
Sciences, 45195-159, Zanjan, Iran\\
$^{2}$ Max-Planck-Institute f\"ur Physik komplexer Systeme,
N\"othnitzer Str. 38, 01187 Dresden, Germany\\
$^{3}$ Departement f\"ur Physik und Astronomie, Klingelbergstr. 82,
4056 Basel, Switzerland}

\date{\today}
\begin{abstract}

We propose a three-terminal spin-valve setup, to determine
experimentally the spin-dependent shot noise, which carries
information on the spin-relaxation processes. Based on a
spin-dependent Boltzmann-Langevin approach, we show that the spin
Fano factor, defined as the spin shot noise to the mean charge
current, strongly depends on the spin-flip scattering rate in the
normal wire. While in the parallel configuration the spin Fano
factor always decreases below its unpolarized value with
increasing spin injection, for the antiparallel case it varies
nonmonotonically. We also show that in contrast to the charge
current Fano factor, which varies appreciable only in the
antiparallel case, the spin Fano factor allows for a more
sensitive determination of the spin-flip scattering rate.

\end{abstract}

\pacs{74.40.+k, 72.25.Rb, 72.25.Ba} \maketitle

The importance of shot noise in mesoscopic systems has been recognized
in the past years as a result of extensive experimental and theoretical
studies of currents fluctuations in a wide variety of hybrid structures
involving normal metals, semiconductors and superconductors
\cite{blanterRev,nazarov:03}. Correlations of current fluctuations at
low temperatures provide unique information about the charge, the
statistics and the scattering of the current carriers. In spintronic
structures
\cite{meservey94,levy94,gijs97,prinz98,johnson:85,Jedema00,zaffalon:03,awschalom02},
in which the transport involves both charge and spin degrees of freedom, the
current fluctuations are expected to contain spin-resolved information
on the conductance process. Consequently spin-polarized current
correlations can be used to extract information about spin-dependent
scattering and spin accumulation in ferromagnet(F)-normal-metal(N)
structures.
\par
Until very recently spin-polarized shot noise has received little
attention. Results of the earlier studies of shot noise in FNF
\cite{bulka99} and FIF \cite{nowak99} systems have been explained
in terms of the well known results of the corresponding
unpolarized systems for two spin directions \cite{blanterRev}.
Tserkovnyak and Brataas have found that shot noise in double
barrier FNF structures, in which the F-terminals have noncollinear
magnetizations, depends on the relative orientation of the
magnetizations of the terminals \cite{tserkovnyak01}. Results of
more recent studies
\cite{mishchenko03,sanchez03,mishchenko04,lamacraft04,BZ04,ZB04}
have revealed that the spin-flip scattering in FNF structures can
change the current correlations strongly, depending on the
polarizations of F-terminals. In Ref.~\onlinecite{ZB04} we have
developed a semiclassical theory of spin-polarized current
fluctuations based on the Boltzmann-Langevin kinetic equation
approach\cite{kadomtsev57,kogan69,nagaev92,sukhorukov99}. It has
been shown that in a multi-terminal diffusive FNF system shot
noise and cross-correlations between currents of different
F-terminals can deviate substantially from the unpolarized values,
depending on spin polarizations of F-terminals and the strength of
the spin-flip scattering in the N-metal . All these studies have
focused on the fluctuations of the charge currents. It is also
interesting to study the fluctuations of the spin currents. Shot
noise of spin-current in absence of charge current was considered
in Ref.  \onlinecite{wang03}. In Ref. \onlinecite{sauret03} it has
been shown that the spin-resolved shot noise of unpolarized
currents can be used to probe the attractive or repulsive
correlations induced by interactions.
\par
In this Letter we study the spin-current shot noise in diffusive
spin-valve systems, in which both charge and spin currents can be
present. To measure the correlations between spin-current
fluctuations we propose to use a three-terminal device in which a
normal diffusive wire is connected through tunnel junctions to
three ferromagnetic terminals of which two have perfect
polarizations pointed antiparallel to each other. We show that
spin shot noise can be determined by measuring the charge shot
noise and the cross correlations between currents through the two
perfectly polarized antiparallel terminals which are connected to
the normal wire by tunnel junctions with different conductances.
The third F-terminal can have arbitrary polarization and is used
to inject a spin accumulation into the normal wire.  The
spin-polarized Boltzmann-Langevin approach is used to calculate
both of charge and spin shot noise. In the presence of spin-flip
scattering these two correlations are distinguished from each
other. We present a detail comparison of charge and spin shot
noise for different polarization of the terminals and the
spin-flip scattering strength in the normal wire.
\par
\begin{figure}[tbh] \centerline{\hbox{\epsfxsize=3.5in
\epsffile{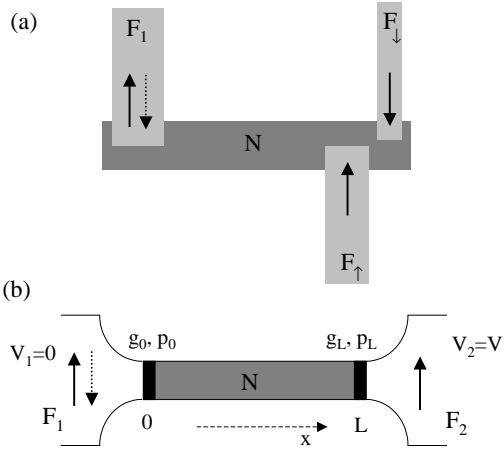}}} \vspace{-2.5in} \caption{(a) Proposed
experimental setup of the three-terminal spin valve to measure
spin current shot noise: a normal diffusive metal strip on which
three ferromagnetic strips are deposited. Two oppositely perfect
polarized ferromagnetic strips are deposited in a distance $L$ of
the third  strip ferromagnet. (b) Schematic of a two terminal
structure equivalent to the three-terminal spin valve.}
\label{zbfig1}
\end{figure}
The layout of the spin-valve system we study is shown in Fig.
\ref{zbfig1}a. Three ferromagnetic terminals F$_1$, F$_{\uparrow}$
and F$_{\downarrow}$ are connected by tunnel junctions to a normal
diffusive wire (N) of length $L$. F$_{\uparrow,\downarrow}$ are
held at the same voltage $V$ and the voltage in F$_1$ is zero. We
model the spin polarization of the terminals as spin-dependent
tunneling conductances of the junctions. The terminals
F$_{\uparrow,\downarrow}$ are perfectly polarized and have
antiparallel polarizations. In this case these two terminals
operate effectively as a single ferromagnetic terminal (held at
the voltage $V$) connected to the wire by a tunnel junction with
conductance given by the sum of the conductances
$g_L=g_{\uparrow}+g_{\downarrow}$ of the two tunnel junctions
connecting F$_{\uparrow,\downarrow}$ to the wire and a
polarization defined as $p_{L}=(g_{\uparrow}-g_{\downarrow})/g_L$.
The junction connecting F$_1$ to the wire has a spin-dependent
conductance $g_{0\alpha}$ ($\alpha=\pm 1$ denotes spin of
electron), which corresponds to the total conductances
$g_{0}=\sum_{\alpha}g_{0\alpha}$ and the polarizations
$p_{0}=\sum_{\alpha}\alpha g_{0\alpha}/g_{0}$. Thus we may
consider the three-terminal structure to be equivalent to a
two-terminal system with corresponding polarizations and the
conductances, as is presented in Fig.~\ref{zbfig1}b.
\par
The requirement that F$_{\uparrow,\downarrow}$ have perfect
polarization is essential for measuring the spin current shot
noise. With this requirement the currents through
F$_{\uparrow,\downarrow}$ are purely polarized. The
cross-correlations $\langle\Delta I_{{\text {c}}\uparrow}(t)
\Delta I_{{\text {c}}\downarrow}(t)\rangle$ between fluctuations
of charge currents through the terminals F$_{\uparrow,\downarrow}$
are simply the correlations between spin up and down currents
$S_{-+}=\langle\Delta I_{-}(t) \Delta I_{+}(t)\rangle$ through the
wire. Thus, measuring the correlations between charge currents
through F$_{\uparrow,\downarrow}$ gives us $S_{-+}$, which can be
used with the correlations of charge current through the wire
$S=\langle\Delta I_{{\text {c}}}(t) \Delta I_{{\text
{c}}}(t)\rangle$ to obtain the spin current correlations
$S_{{\text {s}}}(L)=\langle\Delta I_{{\text {s}}}(L,t) \Delta
I_{{\text {s}}}(L,t)\rangle$ at the point $x=L$. In fact, we have
the relation $S_{{\text {s}}}(L)=S-4S_{-+}$. Here the fluctuations
of charge and spin currents are defined, respectively, as $\Delta
I_{{\text {c}}}(t)=\sum _{\alpha} \Delta I_{\alpha}(L,t)$ and
$\Delta I_{{\text {s}}}(L,t)=\sum _{\alpha}\alpha \Delta
I_{\alpha}(L,t)$ in which $\Delta I_{\alpha}(L,t)$ is the
spin-resolved current fluctuation at the point $L$ of the wire. In
the following, we will calculate the charge and spin current
correlations in the spin-valve system using the corresponding
two-terminal structure shown in Fig. \ref{zbfig1}b.
\par
In the presence of the spin-flip scattering transport of
spin-polarized electrons in the normal wire is described by
Boltzmann-Langevin diffusion equations for the fluctuating charge
and spin current densities at energy $\varepsilon$, $j_{{\text
{c}}({\text {s}})}(x,t,\varepsilon)= {\bar j}_{{\text {c}}({\text
{s}})}(x,\varepsilon) +\delta j_{{\text {c}}({\text
{s}})}(x,t,\varepsilon)$, which read \cite{ZB04}
\begin{eqnarray}
&& \frac{\partial}{\partial x}{j}_{\text {c}}=0, \label{djc}
\\*
&& \frac{\partial}{\partial x}{j}_{\text {s}}=
-\frac{\sigma}{{\ell^{2}_{\text {sf}}}} f_{\text {s}}+i_{\text
{s}}^{{\text {sf}}}, \label{djs}
\\*
&& {j}_{{\text {c}}({\text {s}})}=-\sigma \frac{\partial}{\partial
x}f_{{\text {c}}({\text {s}})}+{j}^{{\text {c}}}_{{\text
{c}}({\text {s}})}. \label{jcs}
\end{eqnarray}
Here $ \sigma= e^2 N_{0} D $ is the conductivity, $ D=
v_{F}^2\tau_{\text {imp}} /3 $ is the diffusion constant, $\ell_{\text
  {sf}}=\sqrt{D\tau_{\text {sf}}}$ is the spin-flip length,
$\tau_{{\text {imp}}({\text {sf}})}$ is the relaxation time of normal
impurity (spin-flip) scattering, $v_{\text {F}}$ is the Fermi velocity
and $N_0$ is the density of states at the Fermi level. The fluctuating
charge and spin distribution functions are expressed as $f_{\text
  {c}}(x,t,\varepsilon)=\sum _{\alpha} f_{\alpha}(x,t,\varepsilon)/2$
and $f_{\text {s}}(x,t,\varepsilon)=\sum _{\alpha} \alpha
f_{\alpha}(x,t,\varepsilon)/2$, respectively, with
$f_{\alpha}(x,t,\varepsilon)$ being the spin-$\alpha$ electron
distribution function. The mean charge and spin distribution functions
obey the equations
\begin{eqnarray}
&& \frac{\partial^2}{\partial x^2}{\bar f}_{{\text {c}}0}=0,
\label{ddfc}\\* && \frac{\partial^2}{\partial x^2}{\bar f}_{\text
{s}0}= \frac{{\bar f}_{{\text s}0}}{{\ell^{2}_{\text {sf}}}}.
\label{ddfs}
\end{eqnarray}
Eqs.~(\ref{djc}-\ref{jcs}) contain the Langevin sources of
fluctuations of the charge (spin) current density ${j}^{c}_{{\text
{c}}({\text {s}})}(x,t,\varepsilon)$ and the divergence term of the
spin current fluctuations $i^{{\text {sf}}}_{{\text
{s}}}(x,t,\varepsilon)$, which reflects the fact that the number of
electrons with specific spin-direction is not conserved by the
spin-flip scattering. The correlators of these fluctuating terms
are given by \cite{ZB04}
\begin{eqnarray}
&& <j_{{\text {c}}({\text {s}})}^{c}(x,t,\varepsilon) j_{{\text
{c}}({\text
{s}})}^{c}(x^{\prime},t^{\prime},\varepsilon^{\prime})>= \Delta
\sigma \sum_{\alpha}\Pi_{\alpha \alpha}(x,\varepsilon),
\label{jsjs}
\\*
&&<j_{{\text {c}}}^{c}(x,t,\varepsilon) j_{\text
{s}}^{c}(x^{\prime}, t^{\prime},\varepsilon^{\prime})>= \Delta
\sigma
 \sum _{\alpha}\alpha\Pi_{\alpha \alpha}(x,\varepsilon),
\label{jsjsp}
\\*
&&<j_{{\text {c}}(\text {s})}^{c}(x,t,\varepsilon) i_{\text
{s}}^{\text {sf}} (x^{\prime},t^{\prime},\varepsilon^{\prime})>
=0, \label{jis}
\\*
&&<i_{\text {s}}^{\text {sf}}(x,t,\varepsilon) i_{\text
{s}}^{\text {sf}}(x^{\prime},t^{\prime},\varepsilon^{\prime})> =
\Delta\frac{\sigma}{2 \ell_{\text {sf}}^2} \sum _{\alpha}
\Pi_{\alpha -\alpha}(x,\varepsilon), \label{isis}
\\*
&&  \nonumber
\end{eqnarray}
where we used the abbreviation $\Delta = \delta(x-x^{\prime})
\delta(t-t^{\prime}) \delta(\varepsilon-\varepsilon^{\prime})$,
and
\begin{equation}
\Pi_{\alpha \alpha^{\prime}}(x,\varepsilon)= {\bar f}_{\alpha
}(x,,\varepsilon) [1-{\bar f}_{\alpha^{\prime}}(x,\varepsilon)].
\label{pi}
\end{equation}
The relations (\ref{jsjs}-\ref{pi}) describe the effect of
spin-polarization and spin-flip scattering on the correlations of
\begin{figure*}
\centerline{\hbox{\epsfxsize=4.5in \epsffile{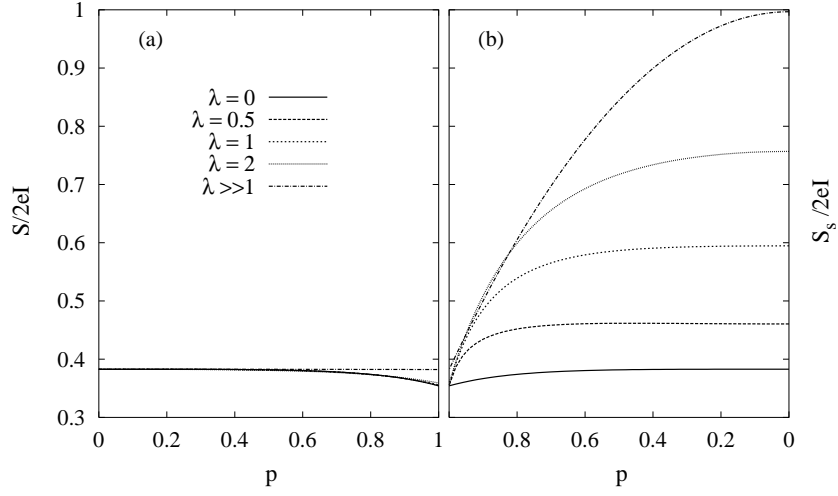} }}
\caption{Charge (a) and spin (b) Fano factors versus the magnitude
of spin polarization of the terminals $p$ for parallel orientation
of the polarizations $p_0=p_L=p$ and when $g_0/g_N=g_L/g_N=1$. The
results are shown for different values of the spin-flip scattering
intensity $\lambda=L/\ell_{\text{sf}}$. }\label{zbfig2}
\end{figure*}
the  current fluctuations sources in the normal wire.
\par
The mean distribution function ${\bar f}_{\alpha 0}={\bar
f}_{{\text {c}}} +\alpha {\bar
 f}_{{\text s}}$ is obtained from the solution of  Eqs.~(\ref{ddfc}) and
(\ref{ddfs}). It reads
\begin{eqnarray}
&& {\bar f}_{\alpha}=f_1+(f_2-f_1)[a+b \frac{x}{L} \nonumber
\\*
&& +\alpha(c \sinh{\frac{\lambda x}{L}}+ d\cosh{\frac{\lambda
x}{L}})], \label{fms}
\end{eqnarray}
where $f_i=f_{\text {FD}}(\varepsilon-eV_i)$ is the Fermi-Dirac
distribution function in the terminal F$_i$ ($i=1,2$) held at
equilibrium in voltage $V_i$ (Fig. \ref{zbfig1}b). The
coefficients $a, b, c, d$ have to be determined from the boundary
conditions, which are  the current conservation rule at the two
connection points of the wire, $x=0,L$.
\par
From the diffusion Eqs.~(\ref{djc}), (\ref{jcs}) and  using Eq.
(\ref{ddfc}) we obtain the expressions for the average and the
fluctuations of the charge current, respectively, as follow
\begin{eqnarray}
&& {\bar I}_{{\text {c}}}(\varepsilon)=b g_{N}(f_2-f_1),
\label{imc}\\* && \Delta I_{{\text {c}}}(\varepsilon)=g_{N}[
\delta f_{\text {c}}(0)-\delta f_{\text {c}}(L)]+\delta I_{\text
{c}}^{c}, \label{delic}
\\*
&& \delta I_{\text {c}}^c(\varepsilon)=\frac{A}{L}\int dx
{j}_{\text {c}}^c, \label{delicc}
\end{eqnarray}
where $g_N=\sigma A/L$ ($A$ being the area of the junctions) is
the conductance of the wire.
\par
Similarly Eqs.  (\ref{djs}), (\ref{jcs}) and (\ref{ddfs}) lead to
the following result for the average and the fluctuations of the spin
currents:
\begin{eqnarray}
&& {\bar I}_{{\text s}}(x,\varepsilon)=g_{N}\lambda (c
\sinh{\frac{\lambda x}{L}}+d\cosh{\frac{\lambda x}{L}})(f_2-f_1),
\label{ims}
\\*
&& \Delta I_{{\text
s}}(0(L),\varepsilon)=\frac{g_N}{s}[-\cosh{\lambda}\delta f_{\text
{s}}(0(L))+\delta f_{\text {s}}(L(0))]
 \nonumber \\*
&& +\delta I^{c}_{\text s}(0(L),\varepsilon), \label{delis}
\\*
&& \delta I^c_{\text s}(0(L),\varepsilon)= A\int  dx (i^{\text
{sf}}_{\text s}+{ j}^c_{\text s}\frac{\partial}{\partial x})
\phi_{{\text s}0(L)}. \label{delics}
\end{eqnarray}
Here $\phi_{{\text s}0} (x)=\sinh{[\lambda(1-x/L)]}/\sinh{\lambda}$,
$\phi_{{\text s}L} (x)=\sinh{(\lambda x/L)}/\sinh{\lambda}$,
$s(\lambda)=\sinh{\lambda}/\lambda$ and
$t(\lambda)=\tanh{\lambda}/\lambda$, and the parameter
$\lambda=L/\ell_{\text {sf}}$ is a measure of the spin-flip scattering.
Note, that as a result of the spin-flip scattering the spin current is
not conserved through the wire.
\par
Following the Boltzmann-Langevin approach, the fluctuating spin
$\alpha$ current at the junctions points $0,L$ can be written as $
I_{\alpha}(0,L)=g_{0(L)\alpha}[f_{0(L)}-f_{\alpha}(0,L)]+\delta
I_{0(L)\alpha}$, in which the intrinsic current fluctuations
$\delta I_{0(L)\alpha}$ is due to the random scattering of
electrons from the tunnel barriers and the fluctuations of the
spin $\alpha$ distribution function are $\delta f_{\alpha}(0,L)$.
From this relation the fluctuating charge and spin currents
through the terminals can be expressed in terms of the fluctuating
spin and charge distributions at the connection points and the
corresponding intrinsic current fluctuations. Denoting $\delta
I_{{\text {c}}({\text s})i}$ as the intrinsic fluctuations of the
charge (spin) current through the tunnel junction, we obtain
\begin{eqnarray}
&&I_{{\text{c}}}(0,L)=g_{0(L)}[f_{0(L)}-f_{{\text
c}}(0,L)]\nonumber \\* && -g_{0(L)}p_{0(L)}f_{{\text
s}}(0,L)+\delta I_{0(L){\text c}}, \label{delic0l}\\*
&&I_{{\text{s}}}(0,L)=g_{0(L)}[f_{0(L)}-f_{{\text
s}}(0,L)]\nonumber \\* &&-g_{0(L)}p_{0(L)}f_{{\text
c}}(0,L)+\delta I_{0(L){\text s}}. \label{delis0l}
\end{eqnarray}
\par
Now we impose the boundary conditions at the junctions. Assuming
spin-conserving tunnel junctions the total (integrated over
energy) spin and charge currents should be conserved at the
junctions points. Using this condition and the expressions for the
spin and charge currents given by Eqs. (\ref{imc}-\ref{delis0l}),
we obtain the coefficients $a, b, c, d$ and the fluctuations of
the charge and spin distributions at $0,L$, $\delta f_{{\text
c}({\text s})}(0,L)$. The results of this calculations can be
inserted in Eqs. (\ref{delic}), (\ref{delis}) to obtain the
fluctuations of the charge and spin currents in the terminal
connected to the point $L$. The results are expressed in terms of
$\delta I_{0(L){\text {c}}}$, $\delta I_{0(L){\text s}}$, $\delta
I_{\text {c}}^c$, and $\delta I^c_{\text s}(0,L)$. To obtain the
current correlations we have to know the correlations between
these terms. The correlations between $\delta I_{\text {c}}^c$,
and $\delta I^c_{\text s}(0,L)$ can be obtained using Eqs.
(\ref{delicc}), (\ref{delics}) and (\ref{jsjs}-\ref{fms}). For
tunnel junctions the correlations of $\delta I_{0(L){\text {c}}}$,
$\delta I_{0(L){\text s}}$ are given by the relations:
\begin{eqnarray}
&&\langle\delta I_{0(L){\text {c}}}\delta I_{0(L){\text
{c}}}\rangle= \langle\delta I_{0(L){\text s}}\delta I_{0(L){\text
s}}\rangle =2e{\bar I}_{{\text {c}}}\label{didi} \\*
&&\langle\delta I_{0(L){\text {c}}}\delta I_{0(L){\text s}}\rangle
=2e(|{\bar I}_{0(L)+}|-|{\bar I}_{0(L)-}|) .\label{didsj}
\end{eqnarray}
where ${\bar I}_{0(L)\alpha}={\bar I}_{0(L){\text {c}}} +\alpha
{\bar I}_{0(L){\text s}}$ is the mean current of spin $\alpha$
electrons.
\par
Using all these results the correlations of the charge and spin
currents $S=\langle\Delta I_{{\text {c}}} \Delta I_{{\text
{c}}}\rangle$ and $S_{{\text {s}}}(L)=\langle\Delta I_{{\text
{s}}}(L) \Delta I_{{\text {c}}}(L)\rangle$ are obtained. In this
way we obtain the charge current Fano factor $F=S/2e|I_{{\text
{c}}}|$ and the spin current Fano factor defined as $F_{{\text
{s}}}=S_{{\text {s}}}/2e|I_{{\text {c}}}|$ in terms of the
normalized conductances $g_{0(L)}/g_N$, the polarizations
$p_{0(L)}$ and the spin-flip strength $\lambda$.
\begin{figure*}
\centerline{\hbox{\epsfxsize=4.5in \epsffile{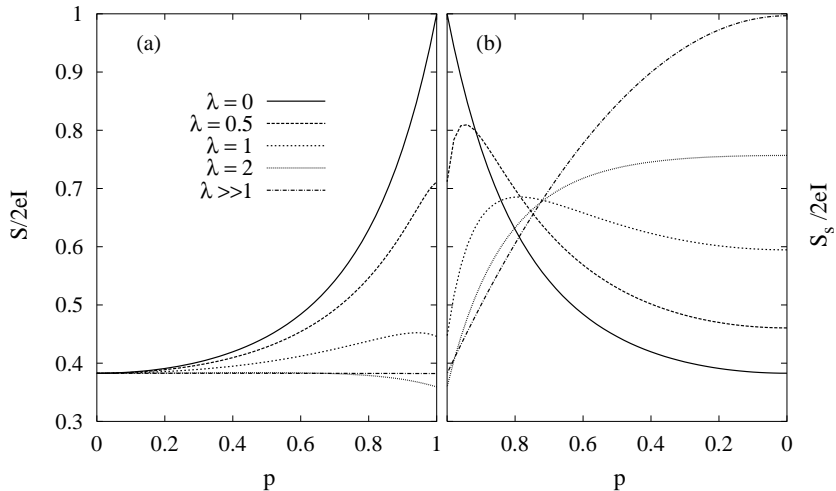} }}
\caption{The same as Fig. \ref{zbfig2} but for antiparallel
configuration of the polarizations $p_0=-p_L=p$.} \label{zbfig3}
\end{figure*}

\par
The final expressions for spin and charge current correlations are too
lengthy to be written here. Figs. \ref{zbfig2} and \ref{zbfig3} show the
dependence of the charge and spin Fano factor on the spin-polarization
of the terminals for different values of the spin-flip scattering
intensity $\lambda$ and when $g_0/g_N=g_L/g_N=1$. Fig.~\ref{zbfig2}
presents the results for the parallel orientation of the magnetizations
of the terminals where $p_0=p_L=p$. In this case $F$
(Fig.~\ref{zbfig2}a) has small variations with respect to the normal
value of $p=0$. For finite $\lambda$ it decreases below the normal value
for $p\sim 1$. For $\lambda\gg 1$ the charge shot noise takes the normal
value for every $p$, which is not surprising since a strong spin-flip
intensity destroys the polarization of the injected electrons from the
terminals. Contrary to charge shot noise, the spin shot noise shows a
strong dependence on $\lambda$ and $p$, as is seen in Fig.
\ref{zbfig2}b. With increasing the polarization $F_{\text s}$ decreases
from its normal value at $p=0$. For small $\lambda$ this variation
occurs only for $p$ close to one. With increasing the spin-flip
intensity the variation is shifted to lower polarizations. At large
$\lambda$, $F_{\text s}$ decreases monotonically from its normal value
$1$ to the minimal value at $p=1$.
\par
The results for the antiparallel configuration $p_0=-p_L=p$ are
shown in Fig. \ref{zbfig3}. For this configuration both, the
charge and spin shot noise, have a strong dependence on the
spin-polarization and spin-flip intensity. At finite $\lambda$ the
charge Fano factor deviates substantially from its normal value
($p=0$) when $p$ increases. The strongest variation occurs when
$\lambda$ tends to zero. $F$ decreases monotonicaly by increasing $p$
and reaches the Poissonian value $1$ at $p=1$. In this case, the
normal wire has perfectly antiparallel polarized leads at its
ends and  constitutes an ideal spin valve, for which the current
vanishes in the limit $\lambda \rightarrow 0$. For very small but
finite $\lambda$ only those of electrons, which undergo spin-flip
scattering once carry a small amount of current. These
spin-flipped electrons are almost uncorrelated and pass through
the normal wire independently resulting in a full Poissonian shot
noise.
\par
The spin Fano factor $F_{\text s}$ has a more complicated dependence on
$p$ for different $\lambda$. While for small $\lambda \ll 1$, $F_{\text
  s}$ increases with increasing $p$ to reach the Poissonian value $1$ at
$p=1$, for large $\lambda \gg 1$ it decreases from the normal value and
reaches a minimum value for a perfect polarization. For $\lambda \sim 1$
the spin shot noise has a nomonotonic behavior with changing the spin
polarization. It is an increasing function of $p$ for small
polarizations and a decreasing function at large polarizations. Thus
$F_{\text s}$ has a maximum value at the polarization which depends on
the spin-flip intensity.
\par
As can be seen in Figs.~\ref{zbfig2} and \ref{zbfig3} the charge and
spin shot noises coincide in two special cases. First, when the
terminals are perfectly polarized and $p=1$, the charge and spin
currents (both mean and fluctuations) are the same and therefore the
correlators of their fluctuations coincide. Second, for a vanishing
spin-flip intensity $\lambda=0$, there is no spin-dependent
scattering mechanism in the whole system (the normal wire and tunnel
junctions). The normal impurity scattering and
the tunnel barriers (assumed to be spin conserving) have the same effect
on the charge and spin transport and the resulting charge and
spin current fluctuations have again the same correlations.
\par
In conclusion we have proposed a three-terminal spin valve structure to
study correlations of spin current fluctuations. The spin valve consists
of a normal diffusive wire which is connected by tunnel contacts to two
oppositely perfect polarized ferromagnetic terminals in one end and to
another ferromagnetic terminals on the other end. Using a spin-dependent
Boltzmann-Langevin approach, the dependence of the spin shot noise on
the spin polarization and the strength of the spin-flip scattering has
been analyzed for parallel and antiparallel configurations of the
polarizations at two ends of the wire. For the parallel case the spin
Fano factor (spin shot noise to the mean charge current ratio) has been
found to decrease with spin polarization from its unpolarized value, but
to increase with the spin-flip rate. In contrast, for the antiparallel
configuration we have found a nonmonotonic behaviour of the spin Fano
factor, depending on the spin-flip scattering rate. We have also found,
that in contrast to the charge Fano factor, which is sensitive to the
spin polarization degree and the spin-flip rate only, in the antiparallel
case, the spin Fano factor shows variations in both configurations. Our
results manifest the effect of competition between spin accumulation and
spin relaxation on the spin current fluctuations in diffusive normal
conductors.

This work was financially supported by the Swiss NSF, the NCCR
Nanoscience, and the RTN Spintronics.

\end{document}